\documentclass[preprint2]{aastex63}

\usepackage{color}

\usepackage[normalem]{ulem}

\usepackage[utf8]{inputenc}
\usepackage{natbib}
\usepackage{graphicx}
\usepackage{longtable}
\usepackage{enumerate}
\usepackage{amsmath}
\usepackage{hyperref}
\usepackage{CJK}
\usepackage[title]{appendix}
\usepackage{rotating}
\usepackage{longtable}
\usepackage{comment}

\def\msun{\ifmmode {\rm M}_{\mathord\odot}\else $M_{\mathord\odot}$\fi}
\def\rsun{\ifmmode {\rm R}_{\mathord\odot}\else $R_{\mathord\odot}$\fi}
\def\lsun{\ifmmode {\rm L}_{\mathord\odot}\else $L_{\mathord\odot}$\fi}

\def\co{$^{12}$CO}
\def\c18o{C$^{18}$O}
\def\h2{H$_{2}$}
\def\hi{{\sc{HI}}}
\def\13co{$^{13}$CO}
\def\n2hp{$_{2}$H$^{+}$}

\def\radmc{{\sc radmc-3d}}
\def\cm2{cm$^{-2}$}

\newcommand{\um}{$\mu$m}

\newcommand{\hii}{H\,{\sc ii}}

\def\um{$\mu$m}
\newcommand{\CASI}{{\sc casi}}
\newcommand{\CASItwoD}{{\sc casi-2d}}
\newcommand{\CASItD}{{\sc casi-3d}}

\def\orion{{\sc orion2}}

\shorttitle{CMR exploration II}
\shortauthors{Xu et al.}

\begin{document}

\title{CMR exploration II -- filament identification with machine learning}

\author{Duo Xu}
\affiliation{Department of Astronomy, University of Virginia, Charlottesville, VA 22904, USA}

\author{Shuo Kong}
\affiliation{Steward Observatory, University of Arizona, Tucson, AZ 85719, USA}

\author{Avichal Kaul}
\affiliation{Steward Observatory, University of Arizona, Tucson, AZ 85719, USA}

\author{H\'ector G. Arce}
\affiliation{Department of Astronomy, Yale University, New Haven, CT 06511, USA}

\author{Volker Ossenkopf-Okada}
\affiliation{I.~Physikalisches Institut, Universit\"at zu K\"oln, Z\"ulpicher Str. 77, D-50937 K\"oln, Germany}

\begin{abstract}
We adopt magnetohydrodynamics (MHD) simulations that model the formation of filamentary molecular clouds via the collision-induced magnetic reconnection (CMR) mechanism under varying physical conditions. We conduct radiative transfer using \radmc\ to generate synthetic dust emission of CMR filaments. We use the previously developed machine learning technique \CASItwoD\ along with the diffusion model to identify the location of CMR filaments in dust emission. Both models showed a high level of accuracy in identifying CMR filaments in the test dataset, with detection rates of over 80\% and 70\%, respectively, at a false detection rate of 5\%. We then apply the models to real Herschel dust observations of different molecular clouds, successfully identifying several high-confidence CMR filament candidates. Notably, the models are able to detect high-confidence CMR filament candidates in Orion A from dust emission, which have previously been identified using molecular line emission.

\end{abstract}


\section{Introduction}\label{sec:intro}

Filaments are an omnipresent structure in the interstellar medium \citep{2014prpl.conf...27A}, and are of great significance in the star formation process, as they contribute to the formation of dense cores and the origin of the initial mass function \citep{2010A&A...518L.102A,2015A&A...584A..91K}. However, the formation mechanism of filaments in molecular clouds remains a subject of debate \citep{Hacar2022}. Observations conducted by the $Herschel$ Space Observatory suggest that filaments have a typical width of 0.1 pc with a notable scatter of $\pm$0.06 pc \citep{2019A&A...621A..42A}, which supports the notion that they may form through the dissipation of large-scale turbulence, occurring at the transitional scale between sonic and subsonic speeds \citep{2011A&A...529L...6A,2020ApJ...904..160F,2021NatAs...5..365F}. Other potential mechanisms for the formation of filamentary structures include the self-gravitational fragmentation of a sheet-like cloud \citep{1983PASJ...35..187T}, the elongation of overdensities by turbulent shear flows, resulting in the development of small line-mass filaments \citep{2013A&A...556A.153H}, the shock compression of magnetized clumps \citep{2021ApJ...916...83A}, the interaction between two shock-compressed sheets \citep{1999ApJ...526..279P}, the convergence of gas flow along local magnetic fields \citep{2014ApJ...785...69C,2015ApJ...810..126C}, the accretion flow driven by gravity \citep{2022MNRAS.512.4715N}, the pressure effects of mechanical and radiation feedback on pre-existing density inhomogeneities on the surfaces and edges of clouds \citep{2019A&A...623A.142S}, the aggregation of small subsonic filaments by collapse and shear flows \citep{2016MNRAS.455.3640S}, and the instabilities induced by large-scale shear and Galactic differential rotation leading to the formation of giant filaments.

Recently, \citet[][hereafter K21]{2021ApJ...906...80K} proposed a novel mechanism for the formation of filaments in molecular clouds, referred to as collision-induced magnetic reconnection (CMR). The study involved magnetohydrodynamics (MHD) simulations of two colliding clouds. The magnetic fields ($B$-fields) in the two molecular clouds are oriented in opposite directions. For example, one cloud has a $B$-field direction pointing towards the positive $z$-axis, while the other cloud has a $B$-field direction pointing towards the negative $z$-axis. The simulations revealed the formation of a stick-like filament at the interface of the two colliding clouds, with a helical magnetic field wrapping around it. This helical magnetic field is known to provide surface magnetic pressure, which helps to confine the filaments and prevent them from dissipating. Zeeman and Faraday rotation measurements have revealed a similar reverse magnetic field at the two sides of the filamentary structures in Orion-A \citep{1997ApJS..111..245H,2019A&A...632A..68T} and Perseus \citep{2022A&A...660A..97T}. This observation is consistent with the formation conditions of CMR filaments. Moreover, the gas morphology of the filaments in the CMR simulations is consistent with that observed in real observations. K21 also noted some unique features of the filaments formed through the CMR mechanism, such as spikes and ring/fork-like structures around the filaments in dust emission. These features are also observed around the filaments in Orion-A, indicating that CMR may be a common mechanism for the formation of filaments.

Identifying filaments formed through the CMR mechanism in observational data poses a significant challenge. The distinctive characteristics of such filaments are difficult to quantify, which makes it challenging to identify them systematically, particularly when dealing with large amounts of observational data obtained from $Herschel$. However, with the help of machine learning, particularly computer vision techniques, it is now possible to systematically identify CMR filaments in dust emission data.

Machine learning has become increasingly popular in astronomy for identifying structures in observational data. Various machine learning algorithms have been widely used for identifying astronomical objects, such as Support Vector Machines (SVM) for protostellar outflows in molecular line emission \citep{2020ApJS..248...15Z}, Random Forests for \hii\ regions in dust emission \citep{2014ApJS..214....3B}, and  Convolutional Neural Networks (CNNs) for segmenting different components of galaxies in multi-optical-band images \citep{2020ApJS..248...20H}. Although SVMs and Random Forests perform well in classification tasks, they require manual extraction of feature vectors from the raw data, which can be challenging and arbitrary. CNNs are a new and powerful approach that can efficiently identify structures or objects in large surveys by adopting raw images or cubes. \citet{2019ApJ...880...83V} developed a Convolutional Approach to Shell/Structure Identification, \CASI, to identify stellar feedback bubbles in 2D density slices and \co\ integrated intensity maps. \citet{2020ApJ...890...64X} and \citet{2020ApJ...905..172X} successfully extended \CASI\ to \CASItD\, which is able to identify stellar feedback bubbles and protostellar outflows in position–position–velocity (PPV) molecular line spectral cubes. \CASItwoD\ and \CASItD\ are able to identify all previously identified bubbles and outflows in nearby molecular clouds in observations, which demonstrates the powerful capability of CNNs to identify features in observational images/spectral cubes. More recently, Denoising Diffusion Probabilistic Models (DDPMs) have demonstrated great proficiency and robustness in image generation \citep{pmlr-v37-sohl-dickstein15,NEURIPS2020_diffusion}. With their well-defined mathematical formulation and ease of training, DDPMs have considered to be particularly well-suited for prediction tasks within the field of astronomy. \citet{2022MNRAS.511.1808S} have demonstrated that DDPMs can generate synthetic observations of galaxies that are highly realistic. Additionally, \citet{2023ApJ...950..146X} have successfully applied DDPMs to infer the mass-weighted gas number density of molecular clouds from their corresponding column density maps and it achieves substantially higher accuracy than that by other methods such as power-law conversion and CNNs.

In this paper, we utilize the previously developed machine learning technique \CASItwoD, in combination with the diffusion model to identify the location of CMR-filaments in dust emission. In Section~\ref{Data and Method} we provide a detailed description of our simulation process and synthetic observations, and  we introduce the \CASItwoD\ algorithm and the diffusion model. In Section~\ref{sec:results} we present the results of our evaluation of \CASItwoD\ and the diffusion model on test data and the prediction they generate on real Herschel dust observations of various molecular clouds. We summarize  our findings and present conclusion in Section~\ref{sec:Conclusions}.

\section{Data and Method}
\label{Data and Method}

\subsection{Numerical Simulations}
\label{Numerical Simulation}
This section introduces two sets of simulations utilized in our study. The first simulation set focuses on modeling the formation of CMR filaments through the collision of two clouds with a reverse magnetic field. These simulations are conducted within a 4 pc box, employing a resolution of 0.0078 pc. The second simulation set involves turbulent boxes and serves as a negative training set, intentionally lacking CMR filaments. These turbulent simulations cover a range of Alfv\'enic Mach numbers, spanning from 0.62 to 1.75. In addition to the aforementioned simulation sets, we introduce a new type of negative training set devoid of CMR filaments but featuring manually created artificial filament-like structures. We provide a comprehensive discussion on these simulated datasets below.

\subsubsection{CMR-Filament Simulations}
\label{CMR-Filament Simulations}

For the CMR-filament models, we use the simulations from \citet[][hereafter K23]{2023ApJS..265...58K}.
K23 conducted a pilot parameter-space exploration\footnote{All simulation data are available at: \url{https://doi.org/10.7910/DVN/CXHWRR}.} for CMR in the context of resistive-MHD. Using the Athena++ code \citep{2020ApJS..249....4S}, K23 explored the effect from seven physical parameters. Specifically, they increased and decreased the fiducial value of each parameter and assessed the impact. In total, there were 14 simulations, including the two with low and high Ohmic resistivity (\mbox{$\eta$\_L}=$1.5\times10^{19}$ cm$^2$ s$^{-1}$ and \mbox{$\eta$\_H}=$1.5\times10^{21}$ cm$^2$ s$^{-1}$), the two with low and high initial magnetic field (\mbox{$B$\_L}=5 $\mu$G and \mbox{$B$\_H}=20 $\mu$G), the two with low and high initial Cloud2 density (\mbox{$\rho_2$\_L} $\rightarrow n_{\rm H_2}=210$ cm$^{-3}$ and \mbox{$\rho_2$\_H} $\rightarrow n_{\rm H_2}=840$ cm$^{-3}$), the two with low and high initial Cloud2 size (\mbox{$R_2$\_L}=0.45 pc and \mbox{$R_2$\_H}=1.8 pc), the two with low and high isothermal temperature (\mbox{$T$\_L}=10 K and \mbox{$T$\_H}=30 K),
the two with low and high collision velocity (\mbox{$v_x$\_L}=0.5 km s$^{-1}$ and \mbox{$v_x$\_H}=2.0 km s$^{-1}$), and the two with low and high shear velocity (\mbox{$v_z$\_L}=0.12 km s$^{-1}$ and \mbox{$v_z$\_H}=0.5 km s$^{-1}$).
Comparing to the fiducial model from K21 (also included in this current study), the K23 explorations formed a variety of CMR-filaments with a range of characteristics. We refer to K23 for details about the simulations. Here we include a few key points about the K23 results.

First, all 14 explorations from K23 form a filament just as the fiducial model from K21. For the case where the two colliding clouds have different sizes and the case where they have different densities, the CMR-filament becomes curved instead of being perfectly straight. 
Second, all the filaments have rich, fiber-like sub-structures. In the synthetic observations, they show up as spikes, rings, and forks, similar to what was seen in K21. These special features set the basis for the machine-learning method to identify CMR-like filaments in large-scale maps. All the 15 CMR simulations (K21 fiducial + K23 14 explorations) are included for machine-learning equally, i.e., each simulation contributes the same number of training images. In principle, there are also kinematic features that could be used to help identify CMR filaments (see K21). However, as we focus here on a comparison to Herschel dust observations where no velocity information is available, we restrict ourselves to the two-dimensional morphological features of CMR-filaments. The kinematics will be considered in a future study.

\subsubsection{Turbulent Cloud Simulations}
\label{Turbulent Cloud Simulations}

To enhance the diversity of the training set, particularly the negative training set that does not include CMR-filaments, we utilize MHD simulations from \citet{2023ApJ...942...95X}.  
The simulations are conducted using \orion\ \citep{2021JOSS....6.3771L} to simulate turbulent clouds with periodic boundary conditions and without self-gravity. Two different mass-to-flux ratios, $\mu_{\Phi}=M_{\rm gas}/M_{\rm \Phi}=2\pi G^{1/2}M_{\rm gas}/(BL^2)$, are adopted: $\mu_{\Phi}$= 1 and 2, which result in an Alfv\'en mach number of 0.6 and 1.8, respectively.

Furthermore, we utilize the MHD simulations presented in \citet{2015ApJ...811..146O}, which aim to investigate the influence of intermediate-mass stellar winds on cloud morphology and turbulence. These simulations consider various magnetic field strengths of 5.6, 13.5, and 30.1~$\mu G$. Further information on the physical parameters of the MHD simulations can be found in the aforementioned studies by \citet{2015ApJ...811..146O} and \citet{2023ApJ...942...95X}. 

It should be emphasized that the MHD simulations conducted by \citet{2023ApJ...942...95X} and \citet{2015ApJ...811..146O} exhibit filamentary structures that are influenced by the magnetic field and formed through mechanisms such as the compression of stellar winds. These filaments, however, are distinct from those formed via the CMR mechanism. These non-CMR filaments are included in the negative training set, allowing the machine learning model to effectively distinguish CMR filaments from other types of filaments.


\subsection{Synthetic Observations}
\label{Synthetic Observations}

\begin{figure}[hbt!]
\centering
\includegraphics[width=0.98\linewidth]{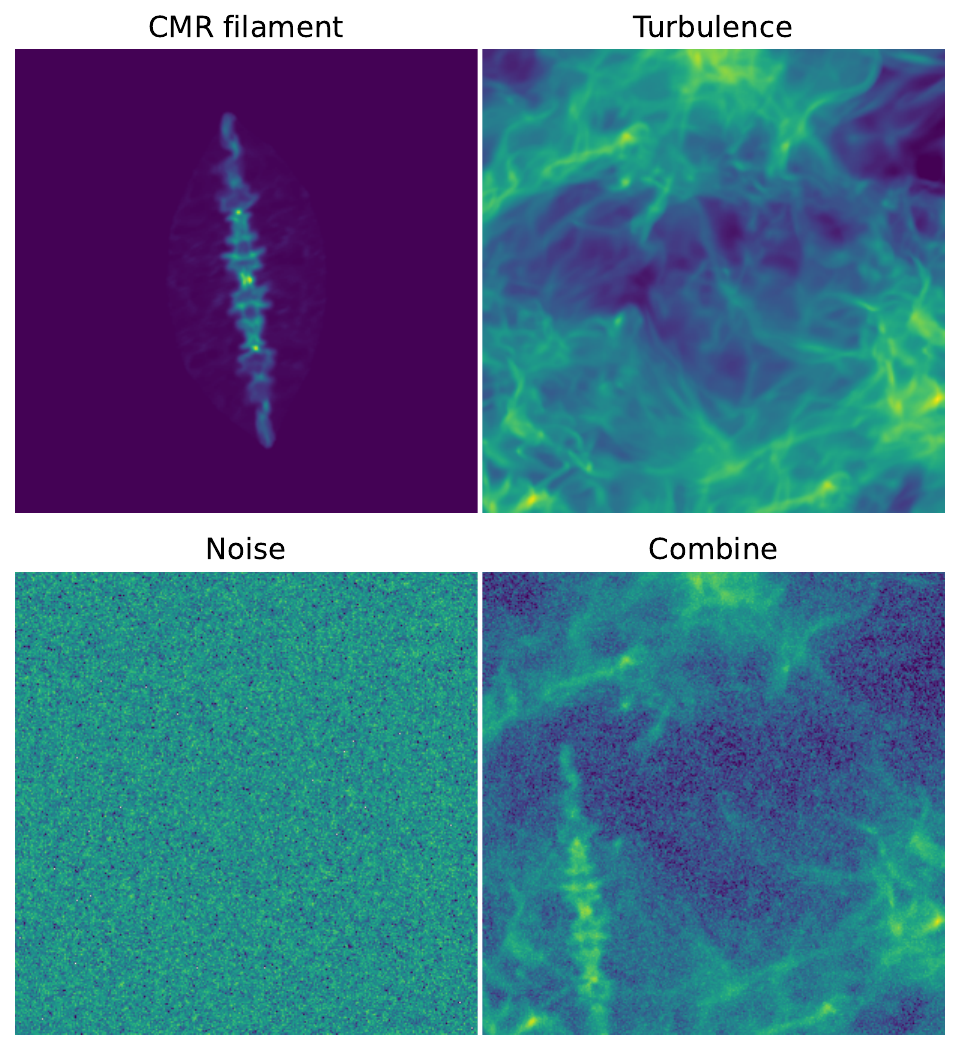}
\caption{An illustration of the process of combining CMR filaments dust emission with turbulent cloud dust emission maps. The upper left panel displays the CMR filament dust emission map, while the upper right panel shows the turbulent cloud dust emission map. The lower left panel represents the Gaussian random noise map, and the lower right panel showcases the combination of these three maps.}
\label{fig.combine_steps_synthetic_1}
\end{figure} 

We employ \radmc\ to perform radiative transfer and generate synthetic dust emission at 250 \um. Our simulation assumes a constant gas-to-dust ratio of 100 and uniform dust temperature of 15 K. Synthetic dust emission is generated from both CMR filaments and turbulent boxes. To simulate real observations where filaments are embedded in complex clouds, we randomly add synthetic CMR filaments emission to the synthetic turbulent images at various orientations and positions. The resulting images represent a composite of both CMR filament and turbulent cloud emission.

An example demonstrating the process of combining CMR filament dust emission into turbulent cloud dust emission maps is presented in Figure~\ref{fig.combine_steps_synthetic_1}. To enhance the diversity of the training set and include a range of brightness levels for CMR filaments in the dust maps, we carefully adjust the contrast when combining them with the turbulent cloud dust emission. The procedure involves several steps. Firstly, the CMR filament map and the turbulent cloud map are scaled to the same intensity scale based on their 97th percentile data value. Next, a scale factor between 0.2 and 1 is randomly chosen and applied to multiply the CMR filament map, which is then added to the turbulent cloud map. The intention behind reducing the intensity of CMR filaments during this addition is to avoid excessive brightness and distinctiveness compared to the turbulent cloud emission. Furthermore, random Gaussian noise is generated with a standard deviation matching that of the turbulent cloud dust map. This noise map is scaled by a random factor between 1 and 3. The resulting noise map is then added to the previously combined maps, yielding the final synthetic dust map that includes both CMR filaments and other filamentary structures with noise.

It is important to acknowledge that in real observations, the distinctive wiggles and spike structures of CMR filaments may be smoothed out by the surrounding dust emission. To address this issue, we have implemented a Gaussian kernel convolution step in our process. The kernel size is randomly selected between 0.01 pc to 0.05 pc, which allows us to generate dust images with relatively smoother features for the CMR filaments. This step is intended to enable the machine learning model to identify CMR filament candidates even with lower resolution, as some candidates may not exhibit pronounced spikes and wiggles. Instead, their intensity and morphology distribution may provide indicators that they are CMR filaments.

It is worth noting that the simulations described in Section~\ref{CMR-Filament Simulations} already include different evolutionary stages, capturing a variety of CMR filament lengths. Additionally, the fiducial box size in our simulations is 4 pc. To further enhance the diversity in terms of physical scale within our training set, we apply a zoom factor of 2 to the synthetic dust images. This enlargement allows the images to encompass a 2 pc scale on each side, effectively covering a wider range of filament lengths. As a result, our synthetic images span a broad spectrum of CMR filament lengths, ranging from sub-pc scale to 4 pc scale. By incorporating these steps, we aim to ensure that our training set exhibits a diverse range of observational scenarios.

Moreover, although we have already included a substantial negative training set consisting of filamentary structures that are not formed via the CMR mechanism, as discussed in Section~\ref{Turbulent Cloud Simulations}, we have also incorporated an additional negative training set. This new set includes randomly generated filaments with a Plummer density profile embedded in the turbulent dust emission. As depicted in Figure~\ref{CMR-Filament Simulations} (and further cases presented in the figures in K23), CMR filaments typically exhibit a straight morphology with distinctive wiggles and spikes perpendicular to their length. In contrast, filaments formed through other mechanisms, such as magnetic field regulation or compression by stellar winds and radiation, exhibit more diverse morphologies, often with curvature.

To simulate these straight CMR filaments, we artificially generate fake straight filaments with a Plummer density profile and integrate them into the turbulent box using the same combination strategy of scaling and adding noise described earlier. It is important to note that both the width and length of these artificial filaments are randomly chosen. Additionally, to mimic the wiggles and spikes observed in CMR filaments, we introduce a random sinusoidal wavy perturbation to the intensity of the artificial stick-like filaments. This perturbation introduces a sinusoidal pattern along the direction of the filaments while maintaining a Plummer density profile perpendicular to the filament direction. An illustrative example can be seen in Sample 6 of Figure~\ref{fig.synthetic-test-casi2d-img}, where Plummer-density-profile filaments with added wavy structures are displayed. These artificially generated straight filaments serve as an additional negative training set to assist the machine learning model in distinguishing CMR filaments from other filamentary structures.

Our data set consists of 16,793 images, including 13,733 combined dust emission images with both CMR filaments and turbulent cloud emission and 3,060 negative training samples which include purely turbulent cloud dust images as well as artificially generated images containing fake filaments. We divide the data set into three subsets for training, validation, and testing, with 60\%, 20\%, and 20\% of images in each subset, respectively.

\subsection{Machine Learning Approaches}
\label{Machine Learning Approaches}

In this section, we introduce two machine learning methods, \CASItwoD\ and the diffusion model to identify the location of CMR-filaments in dust emission.

\subsubsection{\CASItwoD}
\label{CASItwoD}

We employ the same CNN architecture, \CASItwoD, previously developed by \citet{2019ApJ...880...83V}. \CASItwoD\ is an autoencoder that employs residual networks \citep{he2016deep} and a "U-net" \citep{ronneberger2015u}. It consists of two main components: the encoder and decoder parts. The encoder extracts features from the input data and maps them to a lower-dimensional space, referred to as the latent space. The decoder, on the other hand, takes the compressed representation from the latent space and reconstructs the target data. During training, the autoencoder is fed with the combined dust emission map containing both CMR filaments and turbulent cloud emission, and the target data (i.e., the location of CMR filaments). It learns to map the combined dust emission map to the CMR filaments location in a manner that minimizes the difference between the reconstructed output and the target. We utilize the same training hyperparameters as \citet{2020ApJ...890...64X,2020ApJ...905..172X}.

\subsubsection{Denoising Diffusion Probabilistic Models}
\label{Denoising Diffusion Probabilistic Models}

Diffusion models, also known as denoising diffusion probabilistic models (DDPMs) in the deep learning and computer vision research field, are a state-of-the-art generative method that have been used to synthesize high-fidelity data, such as images, videos, and audios \citep{pmlr-v37-sohl-dickstein15,NEURIPS2020_diffusion,stablediff,singer2022make,zhu2022discrete,zhu2023boundary}. The diffusion models are particularly useful for handling complex and high-dimensional datasets with inherent noise and uncertainty. 

The fundamental principles of DDPMs rely on concepts from probability theory and stochastic processes to effectively model and reconstruct data. DDPMs specifically focus on modeling the conditional distribution of clean data given noisy observations. By utilizing probabilistic modeling, DDPMs aim to estimate the underlying distribution of the data, encompassing its statistical properties, patterns, variations, and inherent complexities.

The primary objective of DDPMs is to denoise and reconstruct the original signal from noisy or corrupted observations. By modeling the distribution of clean data and incorporating diffusion processes, DDPMs excel in recovering the true underlying structure while suppressing noise. The key element of DDPMs is the diffusion process, which governs the evolution of the data distribution over time. It initiates from a simple initial distribution and gradually transforms it into the target distribution, which represents the conditional distribution of clean data.

The transformation process in DDPMs occurs through a sequence of diffusion steps. Each diffusion step comprises two primary operations: diffusion and denoising. During the diffusion step, controlled noise is introduced into the data, leading it to evolve along a diffusion path. Subsequently, the denoising step aims to recover the clean data from the noisy observations. This step involves estimating the clean data based on the given noisy observations.

Typically, deep neural network architectures, such as CNNs, are employed in the denoising step. These networks are trained to map the noisy observations to the corresponding clean data. In alignment with our work in this paper, our objective is to reconstruct the filament location based on the complex cloud emission, which encompasses both CMR filaments and turbulent cloud emission.

We adopt the identical configuration of the diffusion model as outlined in the work of \citet{2023ApJ...950..146X}. In their study, a comprehensive mathematical explanation is provided regarding the formulation of DDPMs. To train our diffusion model for the task of recovering the location of CMR filaments from dust emission, we follow the training strategy employed in \citet{2023ApJ...950..146X}.

\subsection{Herschel Gould Belt Survey}
\label{Herschel Gould Belt Survey}

In this section, we introduce the observational data that we used to evaluate the performance of machine learning methods in identifying CMR filaments. For our study, we selected three molecular cloud regions, namely Taurus, Orion, and Auriga–California Cloud. The dust 250 \um\ observations of these three clouds were obtained as part of the Herschel Gould Belt Survey \citep{2010A&A...518L.102A}. The maps were observed using the SPIRE \citep{2010A&A...518L...3G} instrument onboard Herschel \citep{2010A&A...518L...1P}. The clouds were observed with an angular resolution of 18\arcsec.0 at 250 \um. The resulting maps cover an area of 16 deg$^{2}$ for Taurus, 12 deg$^{2}$ for Orion, and 3.4 deg$^{2}$ for the Auriga–California Cloud.

The distance to the Taurus region is estimated to be between 128 pc and 198 pc according to \citet{2019A&A...630A.137G}. To ensure consistency in physical resolution, we downsampled the Taurus Herschel dust map by a factor of 2 to match the resolution of our training set. On the other hand, the distances to the Orion and Auriga–California Cloud regions are approximately 414 pc and 450 pc, respectively, as reported by \citet{2007A&A...474..515M} and \citet{2009ApJ...703...52L}. These distances align well with the physical resolution used in our training set.

\section{Results}\label{sec:results}

\subsection{Test on Synthetic Observations}
\label{Test on Synthetic Observations}

\begin{figure*}[hbt!]
\centering
\includegraphics[width=0.68\linewidth]{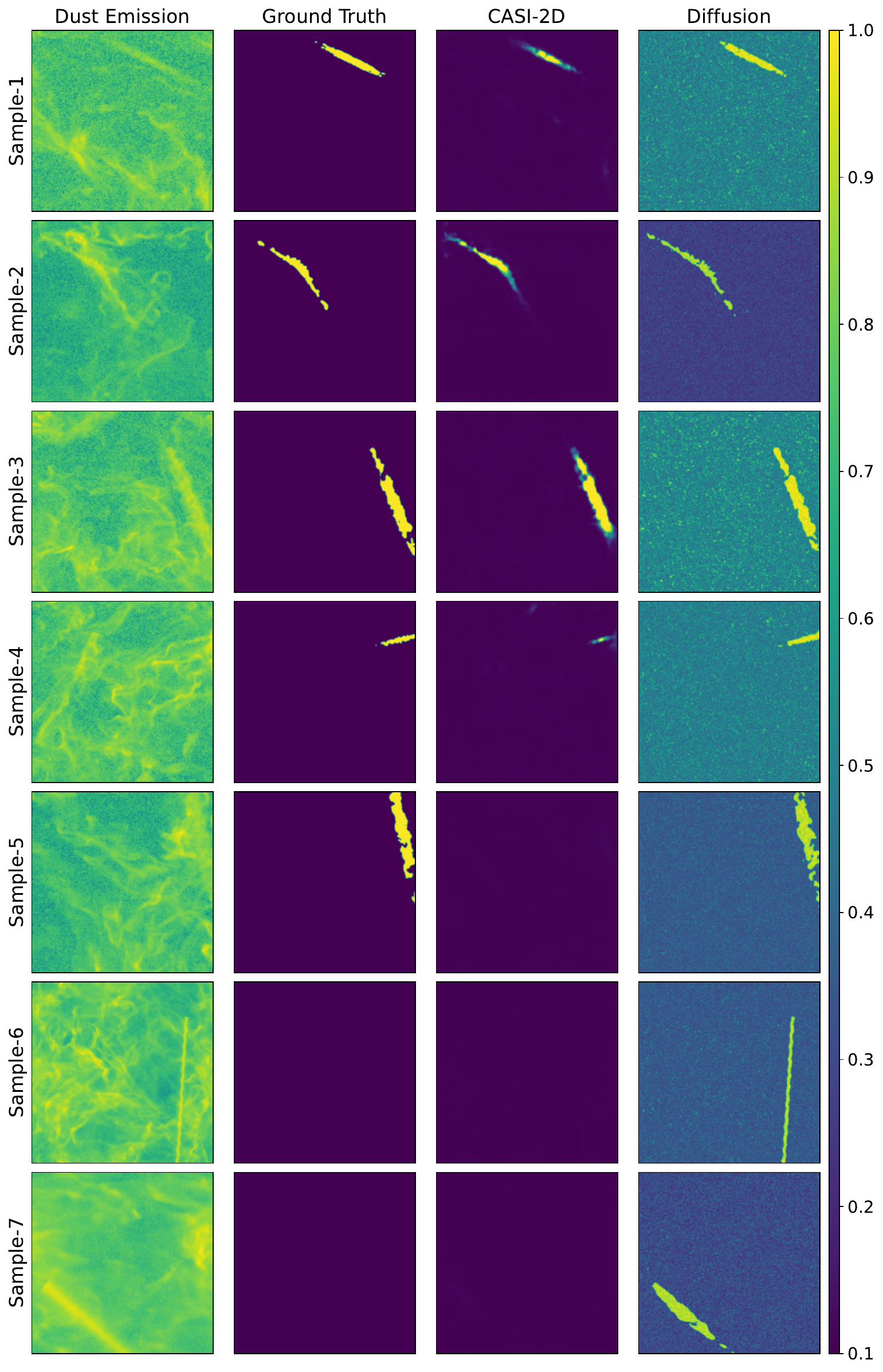}
\caption{Performance of machine learning approaches in identifying CMR filaments. The first column displays the combined dust emission, which includes the emission from both CMR filaments and the turbulent cloud. The second column shows the ground truth, i.e., the actual location of CMR filaments. The CMR filaments are defined by applying a density cutoff ($n>500$ cm$^{-3}$) during the radiative transfer process. This cutoff excludes the low-density ambient gas and focuses solely on the dense structure of CMR filaments. Additionally, we exclude dense structures that are distant from the collision site, as they do not belong to the CMR filaments. The third and fourth columns illustrate the predicted location of CMR filaments by the \CASItwoD\ and the diffusion model, respectively.}
\label{fig.synthetic-test-casi2d-img}
\end{figure*}

In this section, we evaluate the performance of machine learning approaches in identifying CMR filaments in synthetic dust emission. Figure~\ref{fig.synthetic-test-casi2d-img} presents the results of these approaches in detecting CMR filaments on a test set that was not included in our training data. Due to the sensitivity of machine learning models to data with a wide dynamic range, it is necessary to perform data normalization before inputting it into the model. In our case, the input dust emission data, denoted as $x$, is transformed using the equation $x' = \log_{10}(x + 1)$. This transformation helps to reduce the dynamic range of the data and ensures that all values are above zero. Subsequently, we normalize the transformed data, $x'$, to the range between 0 and 1 by dividing $x'$ by its maximum value. The machine learning predictions at each pixel are represented as values between 0 and 1. This is achieved by applying the sigmoid activation function at the last layer of the model.

It is important to note that the morphology of CMR filaments can exhibit variations depending on the initial conditions in the simulation. For instance, in cases where the colliding clouds differ in size or density, the filaments tend to curve rather than remain perfectly straight (Sample-2). Additionally, when the magnetic field strength is weak, the CMR filaments appear shorter compared to other simulations (Sample-4). Furthermore, higher resistivity values result in smoother CMR filaments (Sample-5). Furthermore, the appearance of CMR filaments can be altered by varying viewing angles. 

The dust emission maps in Figure~\ref{fig.synthetic-test-casi2d-img} contain both CMR filaments and other filament structures. In most cases, both \CASItwoD\ and the diffusion model accurately identify only the CMR filaments and exclude other structures with high confidence. However, Figure~\ref{fig.synthetic-test-casi2d-img} also highlights three instances where both models fail. When the CMR filaments are faint and diffuse compared to the host clouds, \CASItwoD\ is unable to detect them, whereas the diffusion model can recover the faint CMR filament (Sample-5). Conversely, if some non-CMR filament
have some features that are similar to some CMR filaments, the diffusion model erroneously identifies it as a CMR filament (Sample-6 and 7). We note that Sample-7, representing the fake Plummer-density-profile filament, exhibits certain similarities to certain CMR filaments (Sample-5). In Sample-5, the smaller structures perpendicular to the CMR filaments appear diffused, and the overall shape appears smooth. This suggests that certain CMR filaments may share a visual resemblance with the smooth Plummer-density-profile filaments. 

From a mathematical perspective, the diffusion model learns the input data distribution and approximates the target distribution by incorporating a Gaussian distribution and Markov chain. In cases where fake stick-like filaments have a data distribution similar to that of CMR filaments, the diffusion model can misclassify them as CMR filaments (e.g., Sample-6 and 7 in Figure~\ref{fig.synthetic-test-casi2d-img}). Conversely, \CASItwoD\ is an autoencoder that utilizes convolutional neural networks to focus on the morphology and structure of the filaments. If these fake stick-like filaments were not part of the positive training set, \CASItwoD\ would not categorize them as CMR filaments. This is also the reason why \CASItwoD\ struggles to identify CMR filaments in scenarios where their emission is faint and diffuse, and the filament structure is not easily recognizable 
(e.g., Sample-5 in Figure~\ref{fig.synthetic-test-casi2d-img}). 

It is of utmost importance to clarify that the concept of the "negative training set" extends beyond images lacking CMR filaments. Even within the synthetic dust map that includes CMR filaments, a notable presence of filamentary-like and fiber-like structures exists. These structures also contribute to the negative training set in our machine learning model, wherein the model learns to distinguish general structures from CMR filament structures. To enhance the visibility of general filaments in the synthetic dust images, we utilize FilFinder \citep{2015MNRAS.452.3435K} to identify and highlight the filamentary structures present in the dust maps. Further elaboration on this can be found in Appendix~\ref{Filamentary Structures Detected by FilFinder}.

\begin{figure*}[hbt!]
\centering
\includegraphics[width=0.48\linewidth]{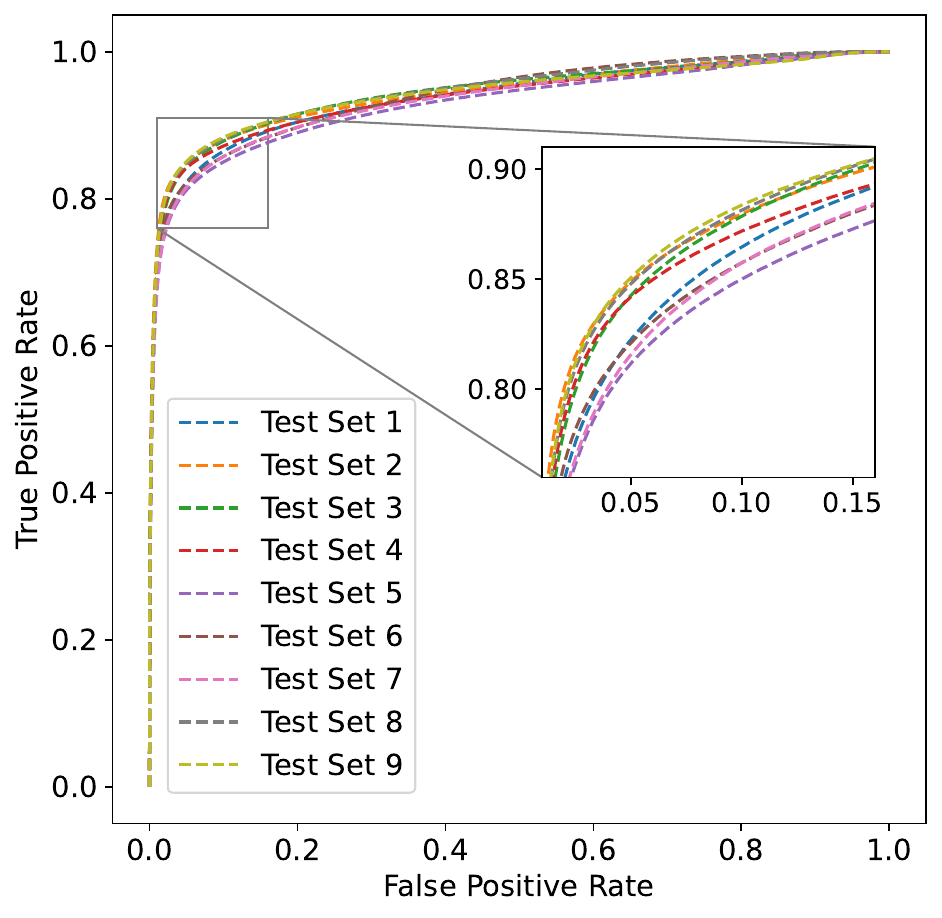}
\includegraphics[width=0.48\linewidth]{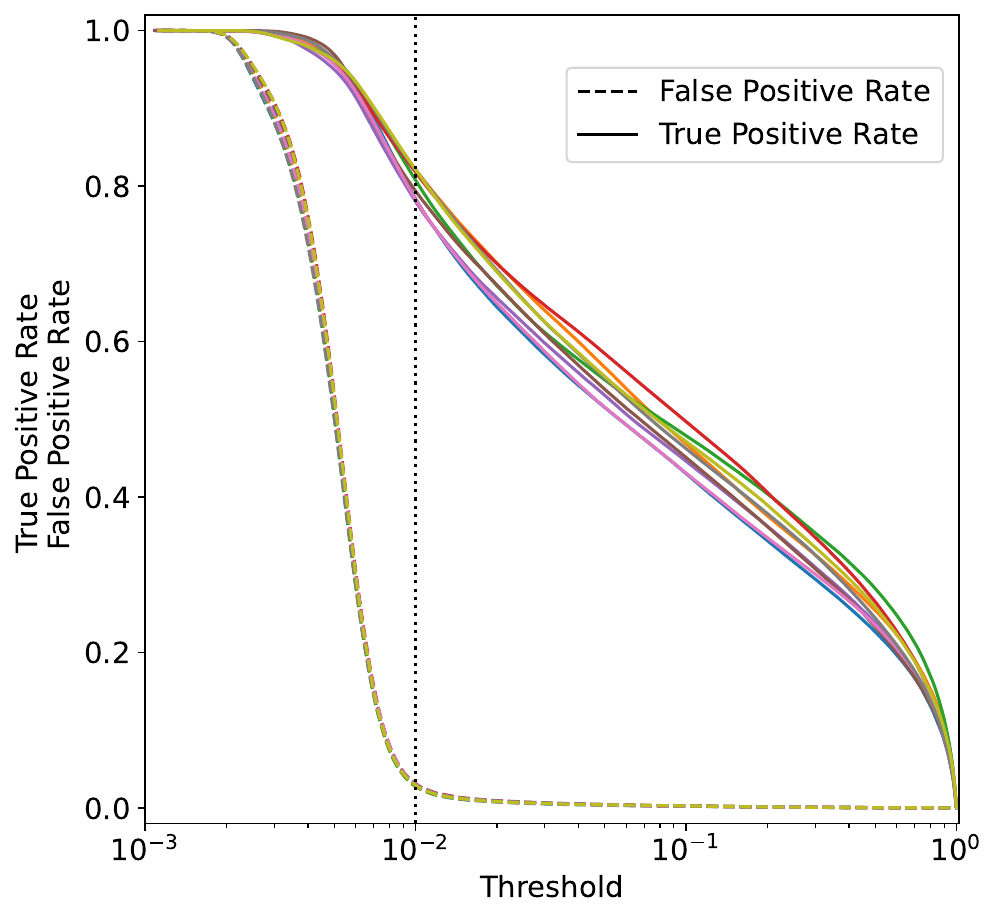}
\caption{The Receiver Operating Characteristic (ROC) curves for the predictions made by \CASItwoD\ (left) and the changes in True Positive Rate (TPR) and False Positive Rate (FPR) as the prediction threshold is adjusted (right). By selecting a threshold of 0.01, a relatively high TPR of 80\% can be achieved while keeping the FPR below 2\%.}
\label{fig.roc_filaments_0304_cnn}
\end{figure*}

\begin{figure*}[hbt!]
\centering
\includegraphics[width=0.48\linewidth]{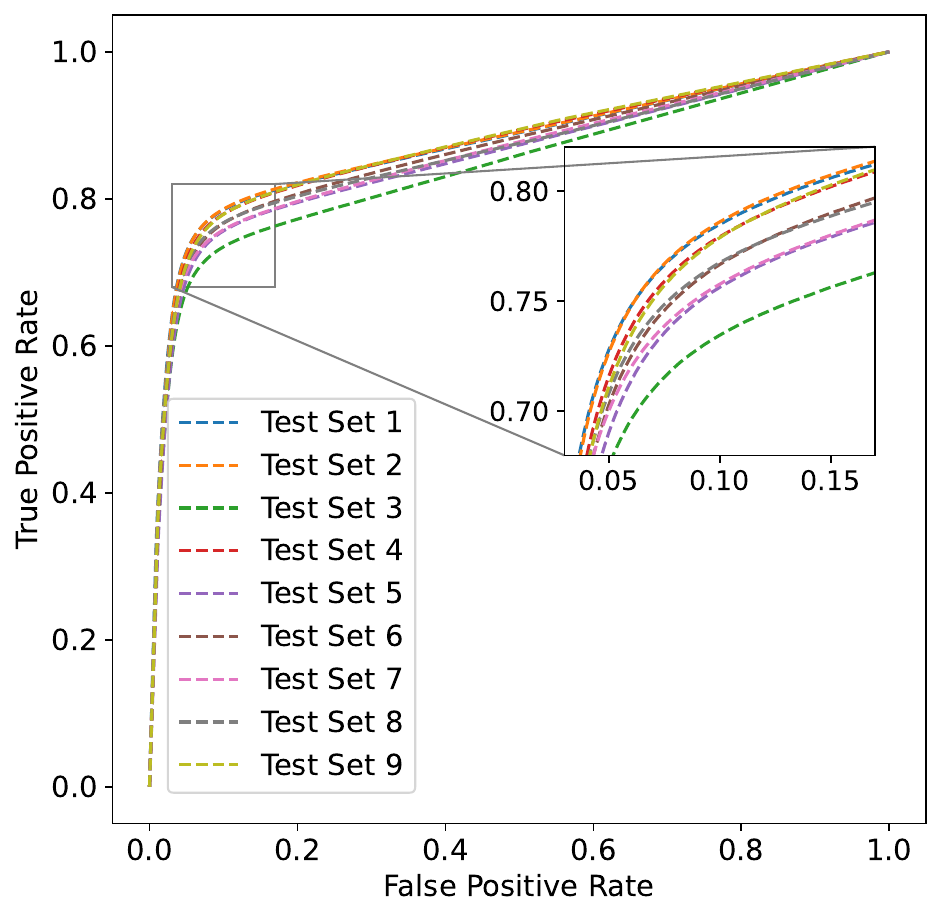}
\includegraphics[width=0.48\linewidth]{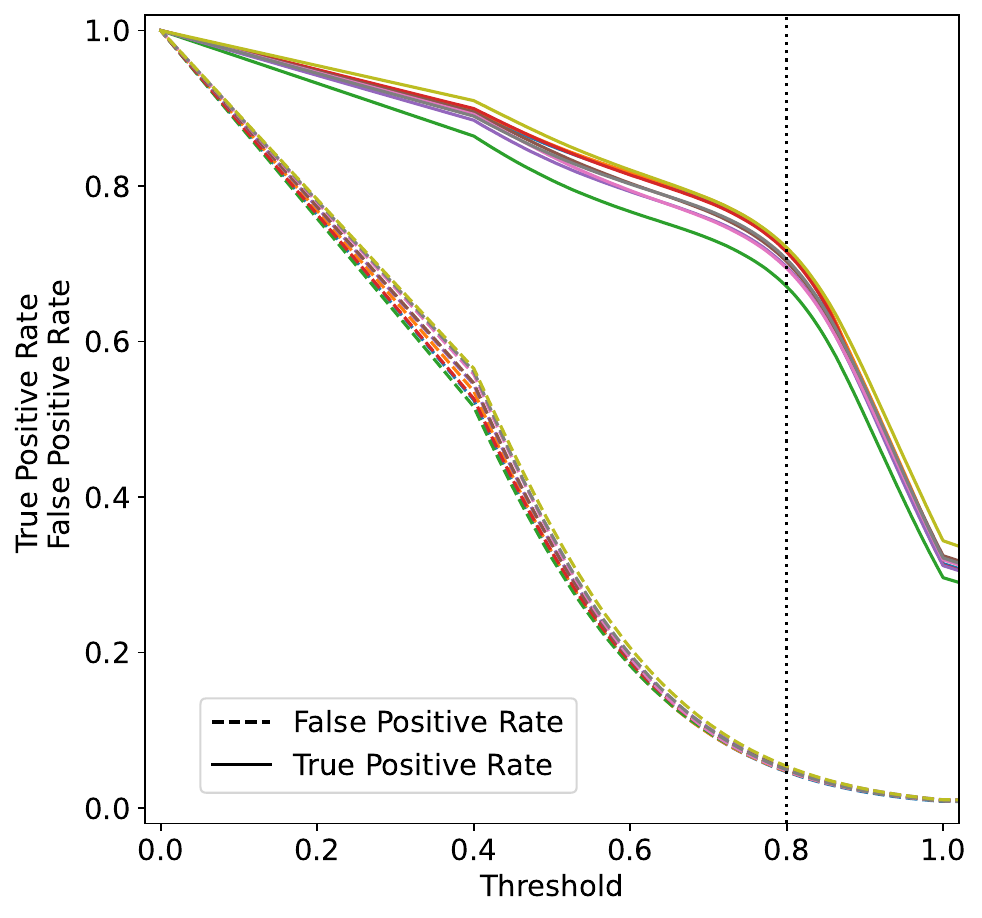}
\caption{The Receiver Operating Characteristic (ROC) curves for the predictions made by the diffusion model (left) and the changes in True Positive Rate (TPR) and False Positive Rate (FPR) as the prediction threshold is adjusted (right). By selecting a threshold of 0.8, a relatively high TPR of $>$70\% can be achieved while keeping the FPR below 5\%.}
\label{fig.roc_filaments_0304_diff}
\end{figure*}

It is important to highlight that the predicted values generated by both machine learning models fall within the range of 0 to 1 for each pixel. However, it is crucial to note that the distribution of predicted values differs significantly between the two models due to their intrinsic differences. In the case of \CASItwoD, which operates as an autoencoder aiming to reconstruct the filament's location, it is highly sensitive to CMR filament-like structures while potentially disregarding other structures. Consequently, the predicted values in non-CMR filament regions can approach zero. On the other hand, the diffusion model's primary focus is on learning noise properties during the denoising process, rather than explicitly capturing the structure or morphology. As a result, the average prediction in non-CMR filament regions does not converge to zero but rather resembles Gaussian noise centered around 0.5. Therefore, it becomes crucial to select an appropriate threshold for the predicted values from both models in order to obtain a mask specifically highlighting the location of CMR filaments.

In order to measure the performance of both machine learning models, we utilized a Receiver Operating Characteristic (ROC) curve, which illustrates the trade-off between the true positive rate (TPR) and the false positive rate (FPR) at various classification thresholds. To generate the ROC curve, different thresholds are applied to the model's predictions. It is crucial to note that the ROC curve is calculated at the pixel level, rather than at the image or filament level. The ground truth values associated with each pixel are labeled as 1 or 0 to indicate the presence or absence of the target class, respectively. The model's predictions associated with each pixel are continuous values ranging from 0 to 1. By setting different cutoff thresholds, such as 0.9, we can determine which samples (represented by individual pixels) are classified as true or false based on the probability threshold. The true positive rate is calculated as the ratio of correctly identified positive pixels to all actual positive pixels, while the false positive rate is the ratio of incorrectly identified negative pixels to all actual negative pixels. By calculating the TPR and FPR for different thresholds, we can construct the ROC curve, where the TPR is plotted on the y-axis and the FPR is plotted on the x-axis. This curve provides valuable insights into the model's performance and allows for the selection of an optimal threshold based on the specific requirements of the classification task. The ROC curve analysis is depicted in Figure~\ref{fig.roc_filaments_0304_cnn} and \ref{fig.roc_filaments_0304_diff}, where the left panels represent the ROC curve for the predictions generated by \CASItwoD\ and the diffusion model, respectively. The right panels display the variations in TPR and FPR as the threshold for prediction is adjusted. In order to minimize bias in the selection of test samples for constructing the ROC curve, we randomly selected test samples from the entire test set a total of 9 times. Each subset of test samples, as indicated in Figures ~\ref{fig.roc_filaments_0304_cnn} and ~\ref{fig.roc_filaments_0304_diff}, consisted of 600 images. Each image had dimensions of 512$\times$512 pixels. Thus, the ROC curve was computed based on a total of 157,286,400 pixels in each subset.

In the right panel of Figure~\ref{fig.roc_filaments_0304_cnn} and ~\ref{fig.roc_filaments_0304_diff}, we can observe the influence of different threshold choices on CMR filament identification. When considering \CASItwoD's prediction, applying a threshold of 0.01 yields a relatively high TPR of 80\% and maintains a low FPR below 2\%. This indicates that by setting a cutoff of 0.01 to mask the \CASItwoD\ prediction, we can recover 80\% of the CMR filament pixels in the test sample while having less than 2\% of the pixels misclassified as CMR filaments. As mentioned earlier, the diffusion model's predicted values follow a different data distribution. By employing a threshold of 0.8 for the diffusion model prediction, we achieve a comparable TPR of over 70\% and a low FPR of 5\%. It is important to note that the choice of this threshold is somewhat arbitrary and depends on our specific scientific objectives. For instance, if we prioritize a cleaner sample with minimal false detections, we can set relatively high thresholds, such as 0.02 for \CASItwoD\ prediction and 0.9 for the diffusion model prediction. However, this might lead to the exclusion of a significant number of true CMR filament pixels. On the other hand, if we prefer a more comprehensive sample and can tolerate a higher false detection rate, we can lower the thresholds, for example, 0.005 for \CASItwoD\ prediction and 0.7 for the diffusion model prediction. To summarize, we suggest using cutoff values ranging from 0.005 to 0.2 for \CASItwoD\ prediction and cutoff values between 0.7 and 0.9 for the diffusion model prediction. We have demonstrated the robustness of these threshold choices when applied to observational data in Figure~\ref{fig.herschel_L1521_250_rot} to \ref{fig.aur_central_250_rot_crop1_hdr}, where the raw predicted values by both models are shown. The predicted values by both models on the observational data align well with those on the synthetic data. 

These ROC curves and corresponding TPR-FPR-Threshold plots provide a comprehensive understanding of the models' performance at different threshold settings, enabling the selection of an optimal threshold based on the desired trade-off between TPR and FPR.

Consequently, we can conclude that the combination of predictions from both \CASItwoD\ and the diffusion model yields highly reliable CMR filament candidates in dust emission. In scenarios where both models identify a structure as CMR filaments and their morphologies align, it is classified as a high-confidence CMR filament candidate. Conversely, if only one model identifies a structure as CMR filaments while the other does not, or if both models agree but their morphologies differ, it is categorized as a low-confidence CMR filament candidate. Different threshold values can be employed in the model predictions to obtain a set of pristine high-confidence CMR filament candidates or a more comprehensive collection with lower confidence levels.

\subsection{Assessing the Performance of Machine Learning Approaches on New Simulations}
\label{Assessing the Performance of Machine Learning Approaches on New Simulations}

\begin{figure*}[hbt!]
\centering
\includegraphics[width=0.8\linewidth]{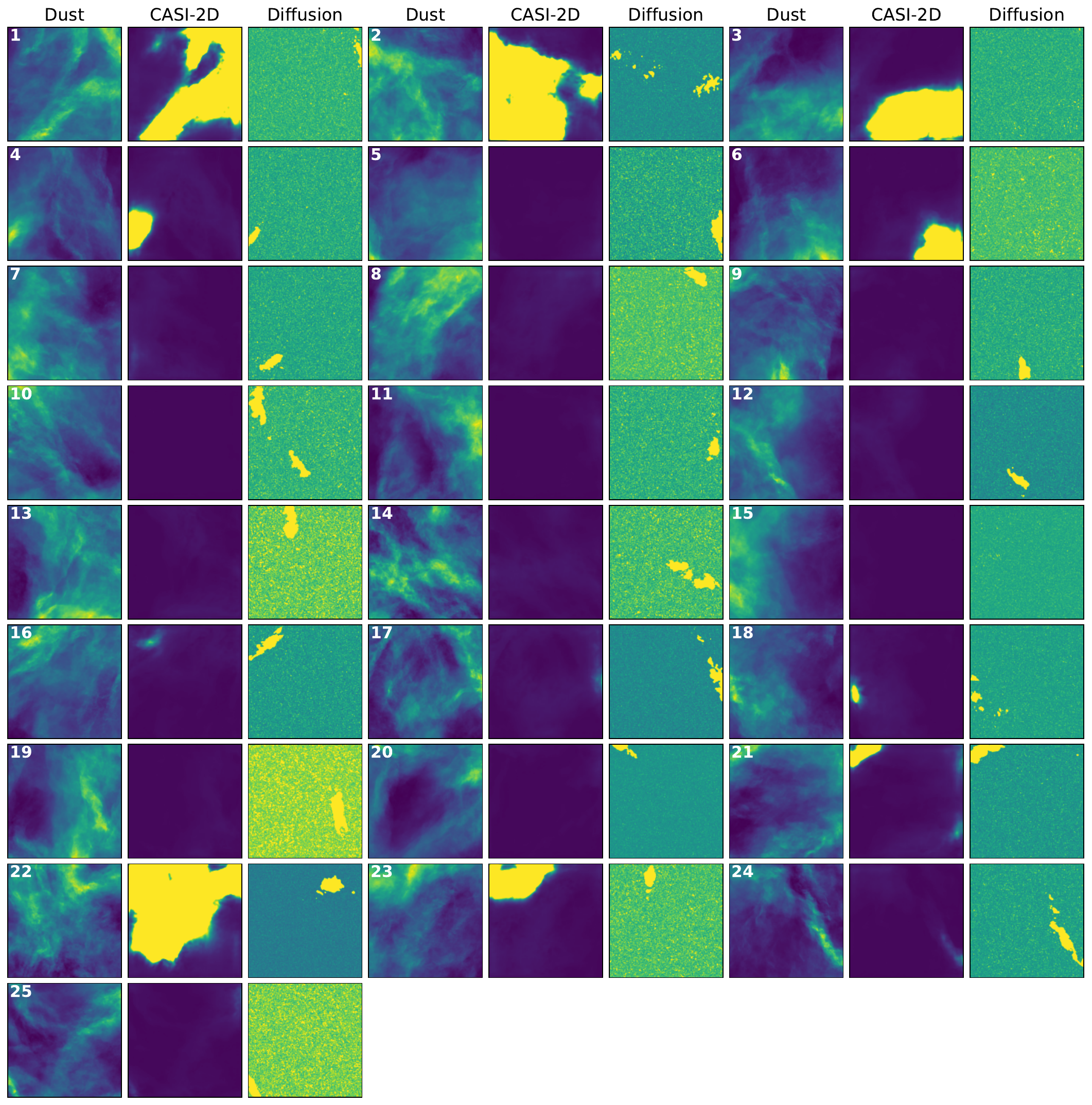}
\caption{Performance of \CASItwoD\ and the diffusion model in identifying CMR filaments on cropped synthetic images that do not contain CMR filaments. }
\label{fig.pred_turbflash_crop_cnn_diffusion}
\end{figure*}

\begin{figure*}[hbt!]
\centering
\includegraphics[width=0.8\linewidth]{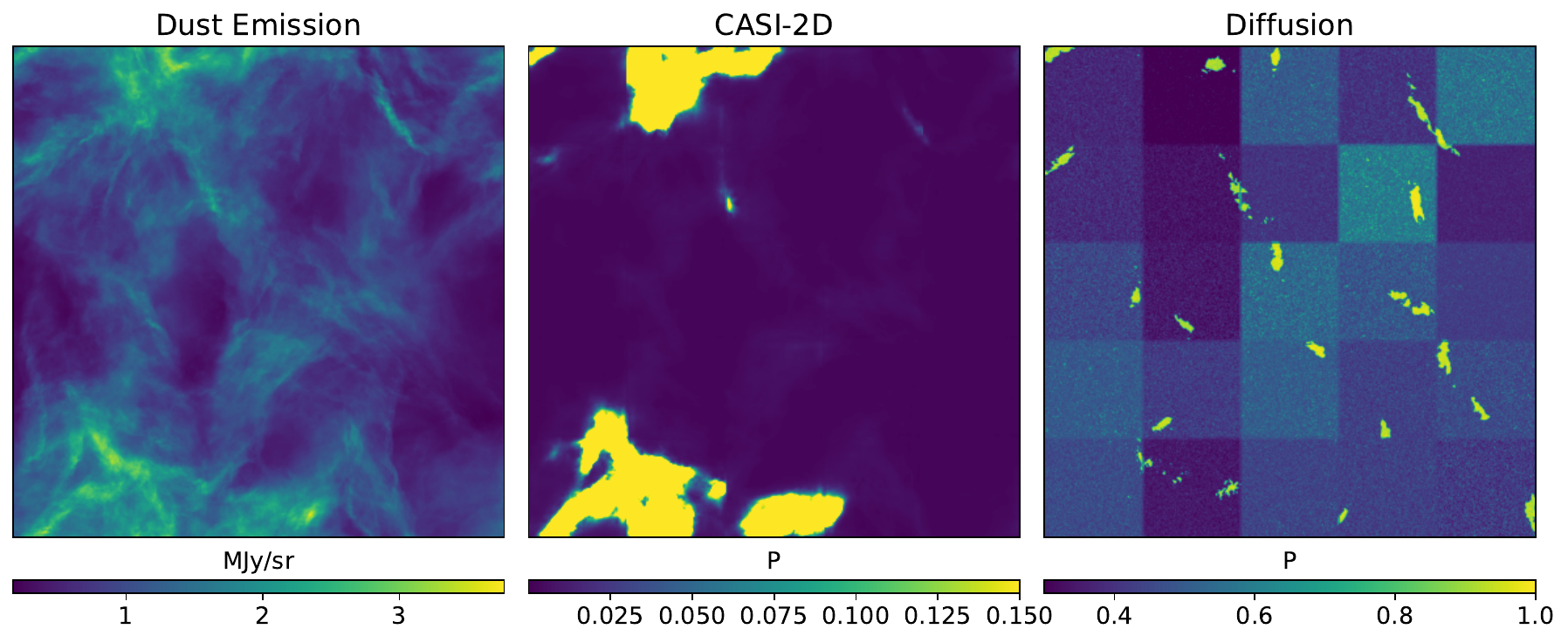}
\caption{Performance of \CASItwoD\ and the diffusion model in identifying CMR filaments on combined synthetic images that do not contain CMR filaments. }
\label{fig.pred_turbflash_combine_cnn_diffusion}
\end{figure*}

In this section, we evaluate the performance of \CASItwoD\ and the diffusion model on new simulation data generated using a different code, a modified version of FLASH \citep{2000ApJS..131..273F}. 
Our test case are the isothermal high-resolution hydrodynamic (HD) simulations of \citet{2021NatAs...5..365F}. Without gravity and cooling the simulations are scale free but when following the scaling used by \citet{2021NatAs...5..365F} for the comparison to thermodynamic models the box has a size of 20~pc. The simulations detected a transition from supersonic to subsonic turbulence at a scale of 0.0125 of the box size, corresponding to 0.25~pc and many filaments with a width spanning a factor of about three around this scale. To feed the two-dimensional comparison, we only used the column density maps from these simulations given by the projection along the $z$ axis. 
As a purely hydrodynamic simulation, this represents a turbulent cloud without the presence of the CMR mechanism. Therefore, it serves as a suitable test to assess the false detection rate of the machine learning models. In an ideal scenario, where the models perform perfectly, we would not expect any detection of CMR filaments in the simulation.

Given the extensive dimensions of the synthetic map in terms of pixel count on both axes and the stipulated image size prerequisite of the machine learning models ($512\times512$), we adopt a technique that entails dividing the expansive map into smaller $512\times512$ postage stamps. Owing to computational constraints, we opt for a larger step size of 250 pixels. Subsequent to the prediction stage, these postage stamps are consolidated through the computation of average predictions for each pixel, culminating in the reconstitution of the original extensive map. Figure~\ref{fig.pred_turbflash_crop_cnn_diffusion} illustrates the performance of \CASItwoD\ and the diffusion model in identifying CMR filaments on these cropped synthetic images that do not contain CMR filaments. We observe a significant number of false detections by \CASItwoD. The diffusion model also exhibits noticeable false detections. However, when we combine the predictions of \CASItwoD and the diffusion model, there are only a few overlaps between the two sets of predictions. When employing the criterion that only structures consistently identified as CMR filament candidates in both the diffusion model prediction and \CASItwoD\ prediction are regarded as high-confidence selections, the following holds: instances where both machine learning models identify a structure as a CMR filament, yet exhibit dissimilar morphology between their predictions, are not deemed high-confidence CMR filament candidates. These criteria, as outlined here, will also be applied in the process of real observation identification, detailed in Section~\ref{Test on Herschel Observations}. Upon application of these criteria in this context, a significant reduction in false detections becomes evident. Consequently, within the array of 25 cropped synthetic images showcased in Figure~\ref{fig.pred_turbflash_crop_cnn_diffusion}, only images labeled 16, 21, and 24 meet the criteria for consideration as CMR filament candidates.

Upon closer examination of the morphology of these CMR filament candidates, we find that they indeed resemble some of the CMR filaments in our training set (e.g., Sample-1 and Sample-4 in Figure~\ref{fig.synthetic-test-casi2d-img}). These filaments are relatively straight with some intensity fluctuations (wiggles) along their length, which is distinct from a Plummer-profile filament. Therefore, we acknowledge that our machine learning models may fail when the morphology of non-CMR filaments resembles that of CMR filaments. We treat all the machine learning model-identified CMR filaments as candidates, and further investigation of their actual formation mechanism requires additional observations such as Zeeman observations to determine the magnetic field direction and ancillary molecular line data to study the kinematics of the host molecular clouds.

Figure~\ref{fig.pred_turbflash_combine_cnn_diffusion} demonstrates the performance of \CASItwoD\ and the diffusion model in identifying CMR filaments on the combined synthetic images that do not contain CMR filaments from Figure~\ref{fig.pred_turbflash_crop_cnn_diffusion}. We observe a patch-like background in the diffusion model prediction, which is a result of the rough crop step when cropping the large maps. This artifact can be reduced by adopting a finer crop step size, as demonstrated in Figure~\ref{fig.herschel_L1521_250_rot}, where no patch-like background is present.


\subsection{Test on Herschel Observations}
\label{Test on Herschel Observations}

\begin{figure*}[hbt!]
\centering
\includegraphics[width=0.65\linewidth]{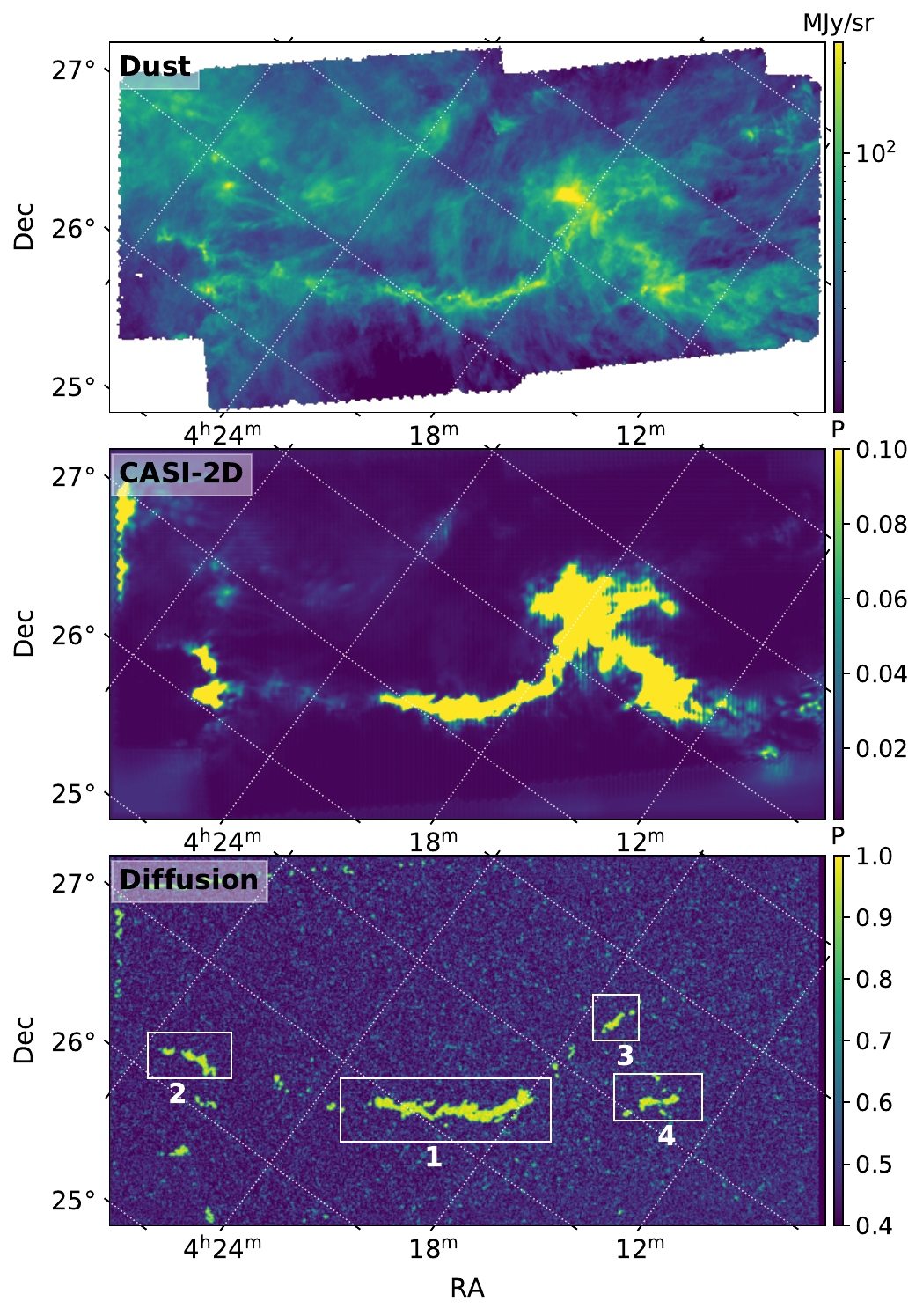}
\caption{Detection of CMR filament candidates in the Taurus B213 region using \CASItwoD\ and the diffusion model. Note that the dust image displayed in this figure has been rotated by a certain degree to minimize the presence of blank regions in the four corners. Consequently, the coordinates in the image no longer align with the x or y axis, but instead follow the trajectory indicated by the white dashed line in the figure. A similar rotation has been applied to Figure~\ref{fig.OrionAN1_250_rot_crop1_raw} and Figure~\ref{fig.aur_central_250_rot_crop1_hdr}.}
\label{fig.herschel_L1521_250_rot}
\end{figure*} 

This section involves the application of both \CASItwoD\ and the diffusion model to real 250\um \/ dust emission data obtained from Herschel observations, which includes the Taurus, Orion, and Auriga–California Cloud regions. Due to the larger size of the Herschel dust map in terms of pixel count on both dimensions and the fixed image size requirement of the machine learning models ($512\times512$), we employ a strategy that involves cropping the large map into smaller postage stamps with dimensions of $512\times512$ and a step size of 16 pixels. After the prediction phase, these postage stamps are combined by averaging the predictions for each pixel, resulting in the reconstruction of the original large map. Figure~\ref{fig.herschel_L1521_250_rot} depicts the outcome of both machine learning models in detecting the CMR filament candidates in the Taurus B213 region. The main B213 filament, located within the white box labeled 1, is proposed as a CMR filament candidate by both \CASItwoD\ and the diffusion model. 

It is important to acknowledge that the predictions made by \CASItwoD\ tend to highlight prominent and bright structures in the Taurus region. This characteristic has been observed before in studies involving the identification of feedback structures in molecular line data, as discussed in \citet{2020ApJ...890...64X,2020ApJ...905..172X}. It is probable that this behavior is a result of the inherent differences in data distribution between the training data and the actual observational data. Conversely, \citet{2023ApJ...950..146X} have demonstrated the stability of the diffusion model when encountering previously unseen data with different data distributions. In Appendix~\ref{Evaluating the Performance of Machine Learning Approaches on Synthetic Images Containing Bright Blobs}, we assess the performance of \CASItwoD\ and the diffusion model on synthetic data that includes bright blobs, leading to an imbalanced data distribution at the tail. Our findings reveal that \CASItwoD\ is significantly influenced by the presence of these bright structures in the images, leading to erroneous identification of the bright blobs as CMR filaments. On the other hand, the diffusion model exhibits greater robustness in the face of high-emission structures due to its ability to learn and incorporate the data distribution. By capturing the statistical properties, patterns, variations, and complexities inherent in the data, the diffusion model transcends reliance solely on morphology for accurate identification. However, as discussed in Appendix~\ref{Evaluating the Performance of Machine Learning Approaches on Synthetic Images Containing Bright Blobs}, it is important to note that in cases where the filamentary structures are significantly dimmer than other structures, there is a possibility that both \CASItwoD\ and the diffusion model may struggle to accurately identify CMR filaments. In such scenarios, both models have a tendency to mistakenly classify bright structures as CMR filaments. The diffusion model prediction in Figure~\ref{fig.herschel_L1521_250_rot} reveals that it focuses on the structure of filaments rather than being concentrated solely on the brightest emission regions, such as the area near box 3. This indicates that the diffusion model is not significantly influenced by the strong emission regions, and its performance is not heavily affected by such regions.

Additionally, it is noteworthy that the structures proposed within boxes labeled 2, 3, and 4 in Figure~\ref{fig.herschel_L1521_250_rot} exhibit different morphologies in the diffusion model's prediction compared to \CASItwoD. To err on the side of caution, we classify the structures proposed within boxes 2, 3, and 4 in Figure~\ref{fig.herschel_L1521_250_rot} as low-confidence candidates.

It is important to note that both machine learning models encounter some challenges in correctly identifying the upper left boundary of the image. Due to the presence of blank regions, which we have filled with zeros to handle NAN numbers, the boundary exhibits a sharp gradient and straight structure. This leads to a misinterpretation by the machine learning models, erroneously identifying the boundary as CMR filaments. When applying machine learning models to real data, we need to exercise caution and be mindful of potential identification issues near the boundaries.


Given that both \CASItwoD\ and the diffusion model accurately propose the Taurus filament in box 1 as exhibiting similar morphology and data distribution to CMR filaments, we strongly suspect that the formation mechanism of this filament is the result of collision between two clouds with opposite magnetic fields. Further Zeeman observations to determine the direction of the magnetic field on both sides of the filaments are crucially needed in order to gain a comprehensive understanding of the forming mechanism of this filament. Most of the Zeeman observations conducted in Taurus have primarily focused on protostellar cores \citep{2000ApJ...537L.139C,2019PASJ...71..117N,2022Natur.601...49C}. However, there are no existing Zeeman observations specifically targeting the two sides of the Taurus B213 filaments to determine the line-of-sight magnetic field direction. If we were able to observe a reversed magnetic field direction on the two sides of the filaments, it could potentially indicate that the filament is formed through the CMR mechanism.

\begin{figure*}[hbt!]
\centering
\includegraphics[width=0.98\linewidth]{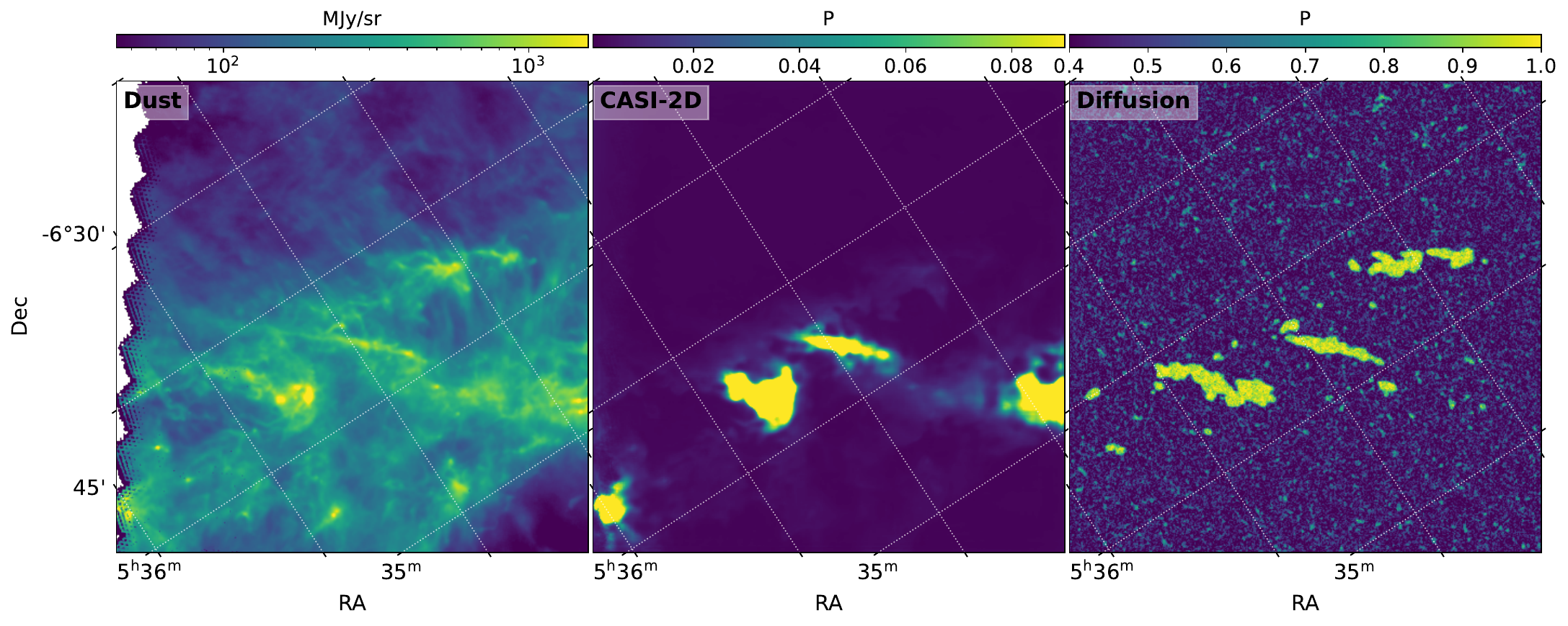}
\caption{Detection of CMR filament candidates in the Orion A north region using \CASItwoD\ and the diffusion model. Note that the stick-like filament in this region has been previously identified as a high-confidence CMR candidate in \citet{2021ApJ...906...80K}. }
\label{fig.OrionAN1_250_rot_crop1_raw}
\end{figure*}

We then evaluated the performance of our two models on the Orion A north region, where the magnetic field directions on the two sides of the cloud are opposite according to \citet{1997ApJS..111..245H}. The stick-like filament in this region has been previously identified as a high-confidence CMR candidate in a detailed study by \citet{2021ApJ...906...80K}. Our models were able to identify the stick-like filaments from the 250 \um\ dust emission, as shown in Figure~\ref{fig.OrionAN1_250_rot_crop1_raw}. It is noteworthy that the initial identification of these filaments was based on molecular line data, which exhibited distinctive rings, wiggles, and fork-like structures. However, in the context of dust emission, these structures tend to appear smoother due to various factors such as integration over the line of sight with different velocity components and the convolutional beam effect in observations. To mitigate this concern, as explained in Section~\ref{Synthetic Observations}, we have incorporated dust images in the training set where the CMR filaments exhibit relatively smoother features due to the convolution with a Gaussian kernel (e.g., Sample-5 in Figure~\ref{fig.synthetic-test-casi2d-img}). This approach enables the machine learning model to identify CMR filament candidates even with lower resolution, as certain candidates may not display distinct spikes and wiggles. Instead, their intensity and morphology distribution can serve as indicators of their CMR filament nature. 

We notice that the diffusion model successfully proposes two additional CMR filaments (lower and upper) alongside the initially proposed CMR filament (middle) in Figure~\ref{fig.OrionAN1_250_rot_crop1_raw}. It is important to highlight that the middle filament exhibits consistent morphology in both the diffusion model prediction and \CASItwoD\ prediction. However, while the lower filament is proposed by \CASItwoD, there is a noticeable difference in morphology between the predictions of the two models for this filament. Therefore, we consider only the middle filament as a high-confidence CMR filament. It is worth noting that the orientations of these three filaments proposed by the diffusion model are somewhat parallel, resembling certain CMR simulations in K23 (e.g., Figure 8), where two filamentary-like substructures form in parallel with the main filament on either side. This suggests that the middle filament in Orion may have formed as a result of the collision between two clouds.

Additionally, we observe that \CASItwoD\ erroneously identifies a bright blob at the right edge of the image as a CMR filament. This is likely due to the presence of bright emission, as discussed earlier, highlighting one of the limitations of \CASItwoD. Crucially, the diffusion model successfully identifies the high-confidence CMR filament (the middle one in Figure~\ref{fig.OrionAN1_250_rot_crop1_raw}) that does not exhibit a markedly strong emission compared to other areas in the dust map. Therefore, we assert that this high-confidence CMR filament is a valid identification and not a consequence of any limitations of the machine learning models. 

It is important to mention that Zeeman observations of \hi\ conducted by \citet{1997ApJS..111..245H} reveal opposite line-of-sight magnetic field directions on the two sides of the Orion filaments. This observation suggests the possibility of two individual clouds with antiparallel magnetic field directions before colliding together to form the filaments. Furthermore, \citet{2021ApJ...906...80K} compared the channel maps and position-velocity diagrams between observations and CMR filament simulations, revealing significant similarities. This provides further evidence that the Orion filaments are likely the result of the CMR mechanism. 



\begin{figure*}[hbt!]
\centering
\includegraphics[width=0.98\linewidth]{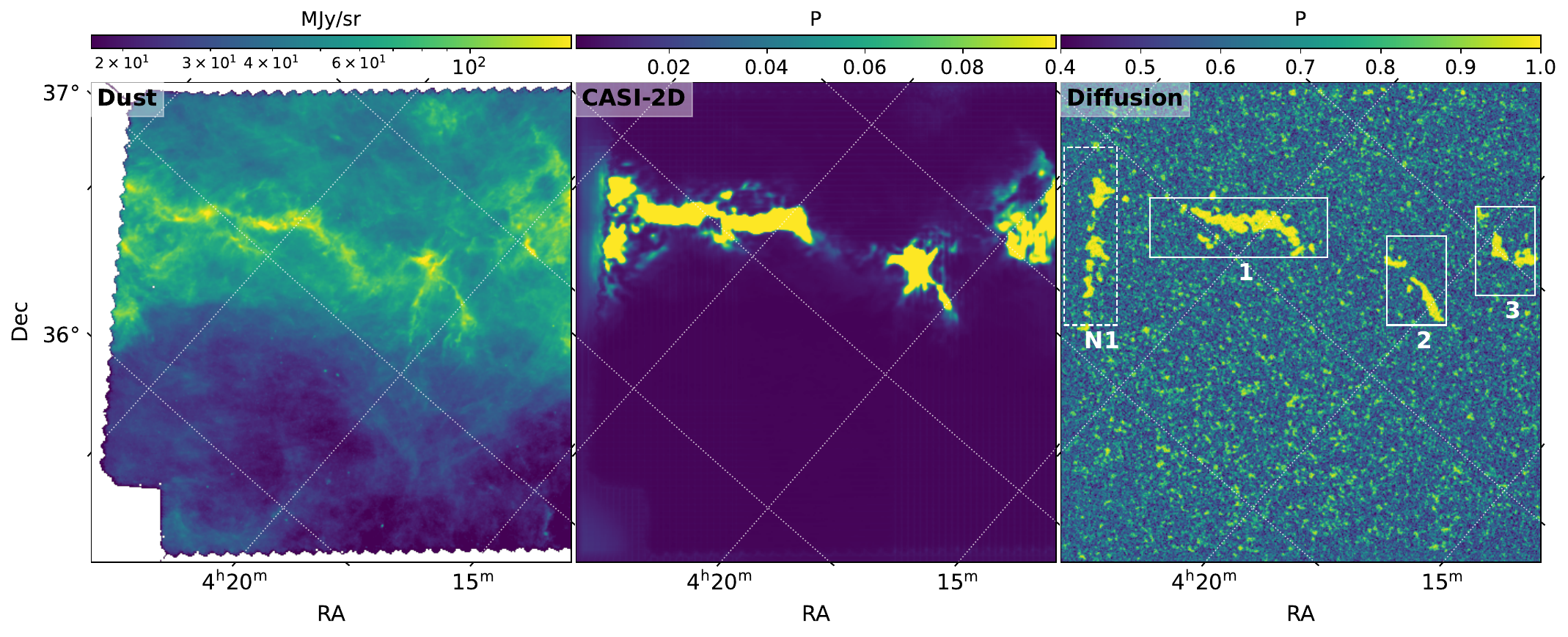}
\caption{Detection of CMR filament candidates in the Auriga–California Cloud region using \CASItwoD\ and the diffusion model.}
\label{fig.aur_central_250_rot_crop1_hdr}
\end{figure*} 

We proceeded to assess the performance of our models on the Auriga-California Cloud and present the results in Figure~\ref{fig.aur_central_250_rot_crop1_hdr}. The predictions of both models exhibit a high level of agreement. Both models correctly propose the filament in box 1 as CMR filaments in Figure~\ref{fig.aur_central_250_rot_crop1_hdr}. Additionally, both models identify two small structures in boxes 2 and 3 in Figure~\ref{fig.aur_central_250_rot_crop1_hdr}. Although the location of the identification in boxes 2 and 3 is similar between the two models, there are differences in the morphology of their predictions. As a result, we consider the two structures proposed in the upper right as low-confidence candidates. Similar to the prediction on the Taurus filament, we notice that both models fail at the boundary of the image, as indicated by box N1 in Figure~\ref{fig.aur_central_250_rot_crop1_hdr}. 

It should be emphasized that the identification of the CMR filament candidates using \CASItwoD\ and the diffusion model is solely based on the morphology and data distribution of dust emission. Therefore, the identifications presented here are only considered candidates. Further investigations on the formation mechanism of these candidates require additional observations, such as Zeeman observations to determine the direction of magnetic fields and ancillary molecular line data to study the kinetics of the host molecular clouds.

\section{Conclusions}\label{sec:Conclusions}

In this paper, we employed the machine learning technique \CASItwoD, in conjunction with the diffusion model, to identify the location of CMR-filaments in dust emission. We evaluated the performance of the two models on test data and applied them to real Herschel dust observations of different molecular clouds. The main results are summarized as follows:

\begin{enumerate}

\item The machine learning techniques, \CASItwoD\ and the diffusion model, demonstrate a high level of accuracy in identifying CMR filaments in the test dataset. By applying a specific threshold to set the false detection rate at 5\%, both models achieve a high detection rate of more than 80\% and 70\% respectively.

\item A limitation of \CASItwoD\ is its potential inability to detect faint CMR filaments in the test set and its susceptibility to misclassifying bright structures as CMR filaments. In contrast, the diffusion model has the potential to misclassify non-CMR filaments that resemble CMR filaments in terms of their data distribution, which includes statistical properties, patterns, variations, and inherent complexities.

\item By combining the predictions of both machine learning models, we can identify high-confidence CMR filament candidates in real Herschel dust observations. Notably, the models detect the high-confidence CMR filament candidates in Orion A from dust emission, which were previously identified using molecular line emission.

\item It is crucial to underscore that the recognition of CMR filaments using both models is designated as high-confidence candidates exclusively. Machine learning techniques never provide an unequivocal guarantee that all identified features perfectly mirror the distinctive characteristics of the training data. Therefore, our identifications represent potential filaments resulting from CMR. For a more comprehensive exploration of the formation mechanism behind these candidates, it becomes imperative to carry out Zeeman observations to determine the direction of magnetic fields, and to analyze accompanying molecular line data to delve into the kinetics of the host molecular clouds.

\end{enumerate}

We would like to express our sincere gratitude for the invaluable suggestions provided by the referee, particularly those related to the machine learning tests conducted on new simulations. These suggestions have indeed played a pivotal role in substantially enhancing the overall quality of the manuscript. D.X. acknowledges support from the Virginia Initiative on Cosmic Origins (VICO). The authors acknowledge Research Computing at The University of Virginia for providing computational resources and technical support that have contributed to the results reported within this publication. We thank the Yale Center for Research Computing for guidance and use of the research computing infrastructure, specifically the Grace cluster. We thank Christoph Federrath for sharing the data of the high resolution turbulence simulations used here as a non-CMR test case for comparison.



\appendix

\section{Filamentary Structures Detected by FilFinder}
\label{Filamentary Structures Detected by FilFinder}

\begin{figure*}[hbt!]
\centering
\includegraphics[width=0.68\linewidth]{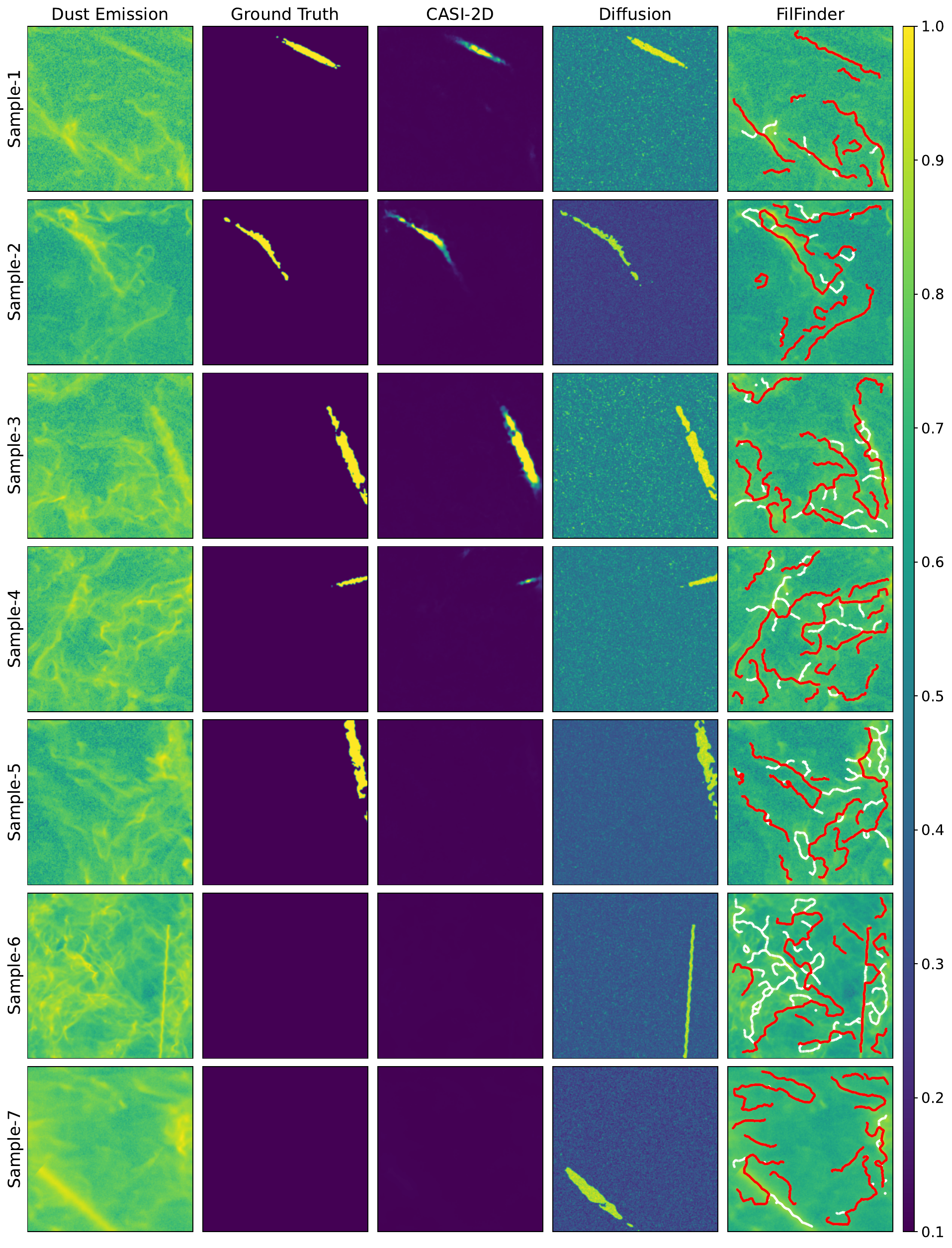}
\caption{Similar to Figure~\ref{fig.synthetic-test-casi2d-img}, but including supplementary panels showcasing the filaments detected by FilFinder.Long filaments are represented by red lines, while short filaments are represented by white lines.}
\label{fig.synthetic-test-casi2d-img-filfinder}
\end{figure*}



In this section, we employ the FilFinder algorithm developed by \citet{2015MNRAS.452.3435K} to analyze the synthetic dust emission data. The purpose is to establish a reference for the presence of filamentary-like and fiber-like structures in the synthetic observations. The parameter configuration for FilFinder is selected in a default manner, devoid of any tuning. Specifically, the parameter values are chosen conservatively. The {\it glob\_thresh}, which represents the minimum intensity required for a pixel to be included in the mask, is set to 5 times the standard deviation of the data (5-sigma) to effectively mitigate noise contamination. The {\it adapt\_thresh} is defined as 0.1 pc, a value corresponding to the typical width of filaments in our Galaxy \citep{2019A&A...621A..42A}. This translates to a width of 12 pixels in the image. The {\it size\_thresh}, indicative of the minimum number of pixels a mask region should encompass to be recognized as genuine, is set to 1000 pixel$^{2}$ to prioritize the identification of substantial filaments rather than shorter fibers. To enhance the visualization of filament locations, FilFinder also implements pruning on branches to eliminate spurious elements. The {\it branch\_thresh}, which sets the minimum length for a branch, is chosen as 40 pixels, while the {\it skel\_thresh}, denoting the minimum length for a skeleton, is set at 120 pixels. These parameters collectively ensure the extraction of a cleaner filament sample with minimal inclusion of spurs.

As illustrated in the fifth column of Figure~\ref{fig.synthetic-test-casi2d-img-filfinder}, it becomes evident that a multitude of filaments and fibers are present in the turbulent simulations. These non-CMR filaments and fibers serve as part of the negative training set, enabling the machine learning models to distinguish between general filaments and CMR filaments. From a quantitative perspective, it is notable that over 75\% of the filaments within a training image are considered negative targets. In our training process, CMR filaments constitute only a small fraction of the overall filamentary structures. Importantly, it should be emphasized that our predictions operate at the pixel level rather than the image level. This signifies that any pixel not associated with CMR filaments is encompassed within the negative training set.

\section{Evaluating the Performance of Machine Learning Approaches on Synthetic Images Containing Bright Blobs}
\label{Evaluating the Performance of Machine Learning Approaches on Synthetic Images Containing Bright Blobs}

\begin{figure*}[hbt!]
\centering
\includegraphics[width=0.8\linewidth]{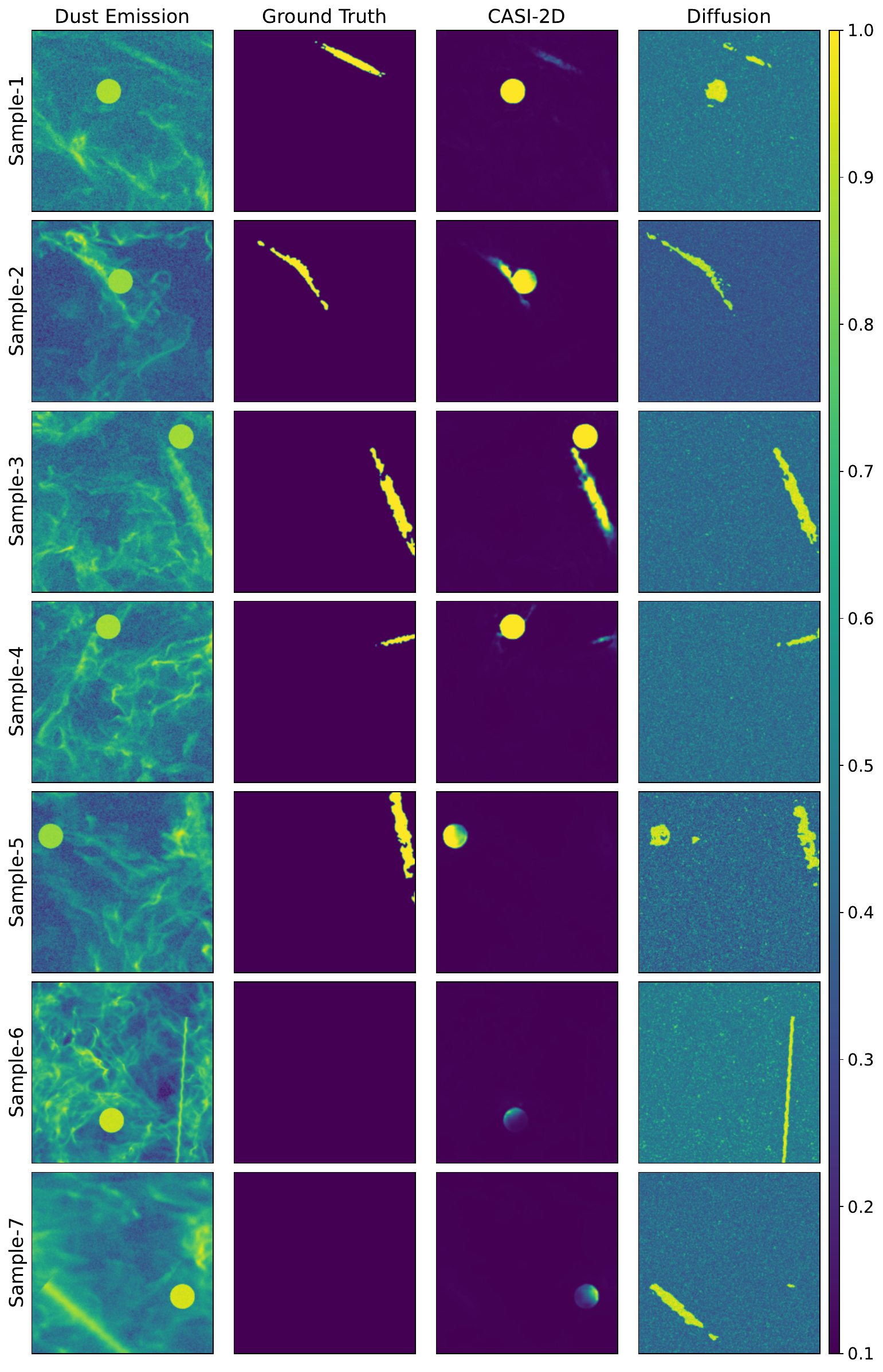}
\caption{Performance of machine learning approaches in identifying CMR filaments on synthetic images containing bright blobs.}
\label{fig.synthetic-test-blob-casi2d-img}
\end{figure*} 

In this section, we assess the performance of \CASItwoD\ and the diffusion model on synthetic images containing bright blobs. To create these images, we introduce a bright blob with a relatively high intensity and add Gaussian random noise within the blob. The intensity of this blob is combined with the synthetic CMR filaments intensity and the turbulent cloud emission, following a similar setup as shown in Figure~\ref{fig.synthetic-test-casi2d-img}.

Figure~\ref{fig.synthetic-test-blob-casi2d-img} presents the performance of \CASItwoD\ and the diffusion model on these images. In the first column, the bright blobs are easily identifiable. However, in the third column, which represents the \CASItwoD\ predictions, we notice that \CASItwoD\ is significantly influenced by these bright structures and fails to recognize them as CMR filaments. In contrast, the diffusion model exhibits greater robustness in handling these unseen images. In most cases, the diffusion model remains unaffected by the presence of these bright blobs in its CMR filament identification. Nevertheless, there are two instances (Sample-1 and 5) where the diffusion model mistakenly identifies the bright blobs as CMR filaments. It appears that when filamentary structures are considerably dimmer than other structures, there is a potential challenge for both \CASItwoD\ and the diffusion model in accurately identifying CMR filaments. Under such circumstances, both models tend to exhibit a tendency to misclassify bright structures as CMR filaments.

It is important to note that such bright blob structures are uncommon in turbulent cloud simulations, indicating that the synthetic images shown in Figure~\ref{fig.synthetic-test-blob-casi2d-img} are novel to the machine learning models. It is natural to expect that the models may encounter challenges with such unseen data. Therefore, the next step in our research will involve incorporating various cloud morphologies into our training set to enhance the adaptability of the machine learning model to different datasets.

\bibliography{ref}
\bibliographystyle{aasjournal}

\end{document}